**THE SUN'S ORIGIN, COMPOSITION AND SOURCE OF ENERGY.** O. Manuel[1], C. Bolon[1], M. Zhong[2] and P. Jangam[3], [1]Nuclear Chemistry, Univ. Missouri, Rolla, MO 65401, om@umr.edu, cbolon@umr.edu, [2]Computational Chem., Iowa State Univ., Ames, IA 50011, [3]Computer Science, Univ. Missouri, Rolla, MO 65401, jangam@umr.edu

**Origin:** The Sun and its planetary system formed from heterogeneous debris of a supernova[1] that exploded 5 Gy ago[2]. Meteorites and planets recorded this as decay products of short-lived nuclides and linked variations in elemental and isotopic abundances[3-11]. Cores of the inner planets grew in a central iron-rich region; the Sun formed on the collapsed SN core.[1,3-11]

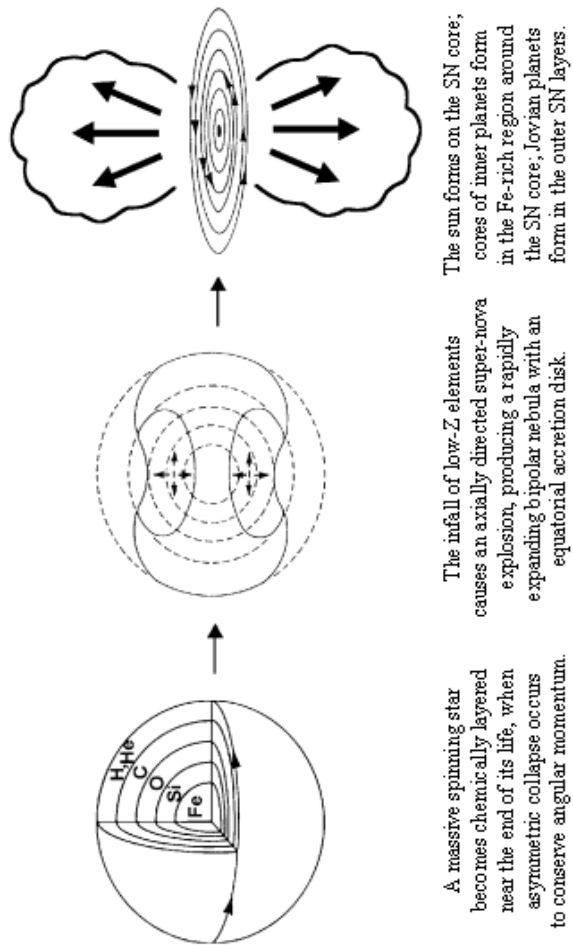

FIG 1. The birth of the solar system from debris of a supernova that exploded here 5 billion years ago.

**Composition:** Diffusion enriches lighter elements and the lighter isotopes of each element at the solar surface[12-14]. When corrected for mass fractionation, the most abundant nuclide that accreted on the Sun[15] was shown to be $^{56}$Fe, the decay product of doubly-magic $^{56}$Ni; the next most abundant nuclide is the doubly-magic $^{16}$O. These nuclides were recently observed[16] in the ash of SN 1987A. The most abundant elements - Fe, Ni, O, Si, S, Mg, and Ca - are the seven, even-Z elements that Harkins[17] found to comprise 99% of ordinary meteorites. The least abundant elements - Li, Be and B - have loosely bound nucleons, confirming a link[17] between abundance and nuclear structure hidden beneath the Sun's H-rich surface, with one conspicuous and important exception - an excess of protons.

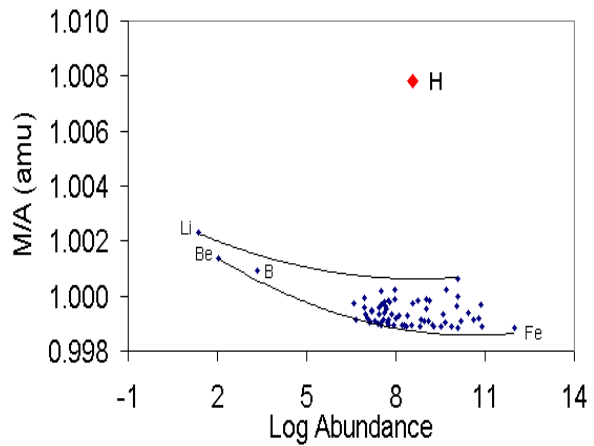

FIG 2. Solar abundance of the elements is related to nuclear stability, as suggested by Harkins[17] in 1917.

**Source of Energy:** 3-D plots[18-20] of energy *vs.* charge density *vs.* mass or atomic number for the ground-state nuclides reveal a cradle, shaped like the trough made by holding two cupped hands together, that contains all nuclear matter in the universe.

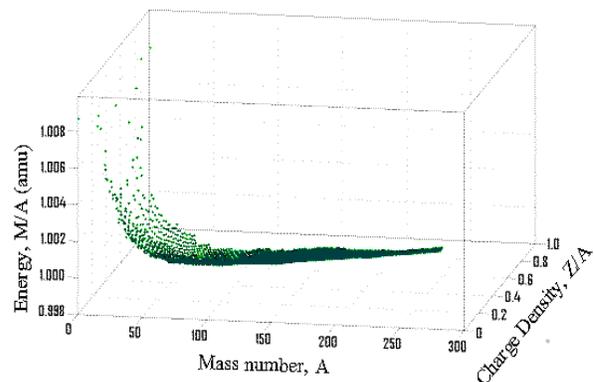

FIG 3. The cradle of nuclear matter in the universe.

The Sun's radiant energy and protons in the solar wind (SW) come from the collapsed supernova core, a neutron star (NS), on which the Sun formed. The universal cradle of nuclear matter (FIG 3) indicates that the energy of each neutron in the Sun's central NS exceeds that of a free neutron by ≈ 10-22 Mev. Solar luminosity and the flux of solar wind protons are generated by a series of reactions: a) escape of neutrons from the central NS, b) decay of free neutrons or their capture by other nuclides, c) fusion and upward migration of H+ through material that accreted on the NS, and d) escape of H+ in the SW. An example might be:

a) The escape of neutrons from the NS,
   $<^1n> \longrightarrow\ ^1n + 10\text{-}22$ Mev

b) The decay of free neutrons,
   $^1n \longrightarrow\ ^1H^+ + e^- + \text{anti-neutrino} + 0.78$ Mev

c) Fusion of hydrogen,
   $4\ ^1H^+ + 2\ e^- \longrightarrow\ ^4He^{++} + 2\ \text{neutrinos} + 26.73$ Mev

d) Escape of excess $H^+$ in the solar wind.

**Implications:** Reactions like a) and b) produce part of the Sun's radiant energy and perhaps the luminosity of isolated neutron stars[21]. Note that reaction a) may release more energy per nucleon than is released by the sum of reactions b) and c), the subsequent decay of the free neutron plus H-fusion. The "Solar Neutrino Puzzle" confirms[22] that reaction c) only generates part of the Sun's total luminosity. Most $^1H^+$ from b) is consumed by H-fusion, but the anomalous abundance of H (Fig. 2) shows that $^1H^+$ also leaks from the interior, selectively carrying lighter nuclides to the solar surface before departing in the solar wind at an emission rate of about $2.7 \times 10^{43}$ $^1H$/yr. Homochirality in living creatures[23] was likely initiated by circularly polarized light (CPL) from the Sun's early NS. Their fate and climate changes of planets[24] may depend on the half-life of this massive nucleus at the Sun's core.

**References:** [1] Manuel O. K. and Sabu D. D. (1975) *Trans. Missouri Acad. Sci., 9*, 104-122; (1977) *Science, 195*, 208-209; (1981) *Geochem. J. 15,* 245-267; (1988) in *Essays in Nuclear, Geo- and Cosmochemistry*, Burgess International, Edina, MN, pp. 1-42. [2] Kuroda P. K. and Myers W. A. (1997) *Radiochim. Acta,* 77, 15-20. [3] Manuel O.K. (1978) *Proc. XII Robert A. Welch Foundation Conf. on Cosmochemistry*, 263-272; (1981) *Geokhimiya*, no. 12, 1776-1800; (2000) in *The Origin of Elements in the Solar System: Implications of Post-1957 Observations*, Kluwer Academic/Plenum Publishers, New York, NY, pp. 607-664. [4] Ballad R. V., Oliver L. L., Downing R. G. and Manuel O. K. (1979), *Nature, 277*, 615-620. [5] Sabu D. D. and Manuel (1980) *Meteoritics, 15*, 117-138. [6] Oliver L. L., Ballad R. V., Richardson J. F. and O. K. Manuel (1981*) J. Inorg. Nucl. Chem., 43*, 2207-2216. [7] Hwaung C.-Y. Golden (1982) *M.S. Thesis*, Univ. Missouri-Rolla, 60 pp. [8] Hwaung G. and Manuel O. K. (1982) *Nature, 299*, 807-810. [9] Manuel O., Windler K., Nolte A., Johannes L., Zirbel J. and Ragland D. (1998*). J. Radioanal. Nucl. Chem., 238*, 119-121. [10] Lee J. T., Li B. and Manuel O. K. (1996) *Geochem. J., 30*, 17-30; (1997*) Comments Astrophys., 18,* 335-345 [11] Manuel O. K., Lee J. T., Ragland D. E., MacElroy J. M. D., Li Bin and Brown Wilbur (1998) *J. Radioanal. Nucl. Chem., 238*, 213-225. [12] Manuel O. K. and Hwaung G.(1983) *LPS XIV*, 458-459; (1983) *Meteoritics, 18*, 209-222. [13] MacElroy J. M. D. and Manuel O. K. (1986) *J. Geophys. Res., 91,* D473-D482. [14] Manuel O. (1986) *LPI Technical Report 86-02, Proc. Workshop on Past and Present Solar Radiation*, p. 28; (1998) *Meteoritics & Planet. Sci., 33*, A97 ; (2000) in *The Origin of Elements in the Solar System: Implications of Post-1957 Observations*, Kluwer Academic/ Plenum Publishers, New York, NY, pp. 285-293. [15] Manuel O. and Bolon C. (2000) "Nuclear Systematics: I. Solar Abundance of the Elements," preprint. [16] Chevalier R. A. (2000) in *The Origin of Elements in the Solar System: Implications of Post-1957 Observations*, Kluwer Academic/ Plenum Publishers, New York, NY, pp. 217-224. [17] Harkins W. D. (1917*) J. Am. Chem. Soc., 39*, 856-879. [18] Manuel O., Bolon C., Zhong M. and Jangam P. (2000*) The Sun's Origin, Composition and Source of Energy*, Report to the Foundation for Chemical Research, Inc., 20pp. [19] Manuel O., Bolon C. and Jangam P. (2000) "Nuclear Systematics: II. The Cradle of the Nuclides," preprint. [20] Manuel O., Bolon C. and Zhong M. (2000) "Nuclear Systematics: III. Solar Luminosity," preprint. [21] van Kerkwijk M. (2000) The Mystery of the Lonely Neutron Star, ESO press release, <http://www.eso.org/outreach/press-rel/pr-2000/pr 19-00.html>. [22] Kirsten T. (1999) *Rev. Mod. Phys., 71,* 1213-1232. [23] Cronin J. R. and Pizzarello S. (1997) *Science, 275,* 951-955 (1997). [24] Malin M. C. and Edgett K. S. (2000) *Science, 290,* 1927-1937.

**Additional Information:** This abstract is condensed from a December 2000 report to the sponsor, the Foundation for Chemical Research, Inc. It will be available on the web at http://www.umr.edu/~om/ with more complete figures and references. Background information is in the Proceedings of the 1999 ACS Symposium on the Origin of Elements in the Solar System: Implications of Post 1956 Observations. For information, contact Susan Safren, Editor, Kluwer Academic/ Plenum Publishers, Susan.Safren@wkap.com. This conclusion to our 40-year effort to understand the origin of the Solar

System and its elements would not have been possible without moral support and encouragement from the late Professors Glenn T. Seaborg and Raymond L. Bisplinghoff.